\documentstyle[pre,preprint,aps]{revtex}
\begin{document}
\draft
\title{Continuum model description of thin film growth morphology}
\author{Chung-Yu Mou$\sp{(a)}$}
\address{Department of Physics, National Tsing-Hua University, \\
Hsinchu, Taiwan 300
, Republic of China}
\author{and}
\author{J. W. P. Hsu$\sp{(b)}$}
\address{Department of Physics, University of Virginia, \\
McCormick Road, Charlottesville, Virginia 22901 USA}
\date{\today }
\maketitle

\begin{abstract}
We examine the applicability of the continuum model to describe
the surface
morphology of a hetero-growth system: compositionally-graded,
relaxed GeSi
films on (001) Si substrates. Surface roughness versus lateral
dimension was
analyzed for samples what were grown under different conditions. We
find
that all samples belong to the same growth class, in which the
surface
roughness scales linearly with lateral size at small scales and
appears to
saturate at large scales. For length scales ranging from 1 nm to 100
$\mu$m,
the scaling behavior can be described by a linear continuum model
consisting
of a surface diffusion term and a Laplacian term. However, in-depth
analysis
on non-universal amplitudes indicates the breaking of up-down
symmetry,
suggesting the presence of non-linear terms in the microscopic model.
We
argue that the leading non-linear term has the form of $\lambda
_1({\bf
\nabla }h)^2$, but its effect on scaling exponents will not be
evident for
length scales less than 1 mm. Therefore, 
the growth dynamics of this system is
described by the Kuramoto-Sivashinsky equation, consisting of the two
linear
terms plus $\lambda _1({\bf \nabla }h)^2$, driven by a Gaussian
noise. We
also discuss the negative coefficient in the Laplacian term as an
instability mechanism responsible for large scale film morphology on
the
final surface.
\end{abstract}


\section{Introduction}

The dynamics of film growth has proven to be rich and interesting
phenomena.
\cite{barabasi} In an ideal homo-growth, during which the deposition
rate is
sufficiently low and the temperature of the substrate is high enough,
the
adatoms have enough time to find their optimal positions so that most
adatoms are registered and the growing front has only small
fluctuations
around the equilibrium shape. The resulting film under this growth
condition
is smooth. Such quality of the film can be maintained indefinitely
only if
the atoms below the growing front are always kept in true
equilibrium. In
reality, however, growth usually happens in non-equilibrium
conditions. In
fact, the real power of thin film growth is the capability to create
new
materials and to obtain desired physical properties via
non-equilibrium
growth. In practical applications, an often encountered situation is
to have
the thin film and the substrate be different materials, i.e.,
hetero-growth.
In hetero-growth, the growth mode is usually not layer-by-layer;\cite
{zangwill} instead, it depends on equilibrium material properties as
well
as kinetic parameters during growth. In the extreme case, it was
demonstrated recently that the coherent strain in the film can be
utilized
to fabricate novel nanostructures. \cite{nanodot}

During the last decade, much work has been devoted to understanding
the
non-equilibrium film growth. There are several important features
observed
in the final surface morphology. First, surfaces are highly
irregular. It is
therefore impractical to predict or describe such surfaces in
microscopic
details. A coarse-grained, statistical modeling is more appropriate.
Secondly, certain large-scale features can survive in the final
surface
morphology. These features are manifestations of underlying
microscopic
instabilities. Thirdly, it is found, in numerous experiments and
computer
simulations, that analysis of the surface roughness versus the
sample's
linear dimensions provides a useful classification of growth
mechanism.\cite
{barabasi,expts} Specifically, many surfaces exhibit self-affinity in
which
a scaling phenomenon is found
\begin{equation}
\sigma (t)\equiv \sqrt{\langle \,[\,H({\bf r},t)-\langle H({\bf
r},t)\rangle
\,]^2\,\rangle }\sim A_\sigma L^\chi f\left[ L/\xi (t)\right] .
\label{scaling}
\end{equation}
Here the sample dimension is $L\times L$, $\sigma (t)$ is the surface
roughness, $H({\bf r},t)$ is the height of the surface at position
${\bf r}$
and time $t$, and $\langle H({\bf r},t) \rangle $ 
is the average height.
The length $\xi (t)$ denotes the
characteristic
length of the surface. If no other important length scale is present,
$\xi
(t)$ is the correlation length built up during the course of film
growth and
scales as $\xi (t)\sim (\tilde{\nu}t)^{\frac 1z}$. The exponents
$\chi $ and
$z$ are useful in characterizing the surface morphology. Different
growth
mechanisms result in different exponents, while the details of growth
manifest themselves only through non-universal amplitudes $A_\sigma $
and $
\tilde{\nu}$. Along with the discovery of the above scaling
phenomenon, much
theoretical work has been devoted to constructing appropriate
continuum
models for describing the film growth. The goal is to reproduce these
scaling results.

Experimentally, until now, most work has concentrated on the
measurement of
the exponents, which leaves open many important issues. For example,
in
previous work, no differentiation in material nature between film and
substrate and no examination of the applicability of continuum models
to
surfaces with large features were ever carefully made. Many results
are thus
{\it a priori} only applicable to homo-growth systems and to growth
without
instability. Results from model systems\cite{expts} show that the
quantitative scaling characteristics of surface roughness depend on
the
particular material system; for example, the data for $\chi $ range
from 0.2
to 1. It is therefore particularly important to examine the
applicability of
theoretical ideas, such as continuum modeling, beyond the homo-growth
system.

In this paper, we examine a hetero-growth system:
compositionally-graded,
relaxed GeSi films on (001) Si substrates. A distinct feature of
surfaces in
this system is the existence of large-scale patterns, known as
cross-hatches.
\cite{hsu} These patterns are closely related the underlying misfit
dislocation network. Much work has been devoted to understanding the
mechanism responsible for the cross-hatch
formation.\cite{fitzgerald,xhatch}
Here we approach this problem differently by examining how
surface
roughness depends on lateral size, i.e., by a scaling analyses. In
addition,
we go beyond the usual approach, that is based solely on the
measurement of
roughness exponents, and perform more comprehensive analyses on
non-universal amplitudes and the up-down symmetry of surfaces. Our
results
indicate that, up to a length scale of 100 $\mu {\rm m}$, a continuum
model
in which the linear parts are composed of a surface diffusion term
and a
Laplacian term is appropriate for describing these surfaces. The
breaking of
up-down symmetry shows that non-linearities must also be present.
Detailed
analysis sets a lower bound of $\approx $ 1 mm for observing the
scaling
exponents that arise from the lowest-order non-linear term. 
The resulting continuum model that is consistent with the
experimental
data is a two-dimensional Kuramoto-Sivashinsky equation driven by a
Gaussian
noise. \cite{KSeq} We also discuss the instability that might be
responsible
for the cross-hatch formation in the framework of a continuum
description.

This paper is organized as follows. Section II briefly reviews
relevant
theoretical ideas and results. In Section III, we apply scaling
analyses to
study the surface morphology of relaxed GeSi films grown on Si
substrates.
Both universal scaling exponents and non-universal amplitudes are
analyzed.
The restrictions placed on the proposed continuum model by the
experimental
results and the crossover from this model to other models are
discussed in
Section IV.
The Appendixes are devoted to more technical details. In Appendix A,
we
calculate the surface roughness for a sinusoidal surface. In Appendix
B, we
apply the Dyson-Wyld renormalized perturbation theory to analyze one
of the
possible growth models, that can account for our results.

\section{Continuum models and theoretical results}

We begin with a brief review of the theoretical situations. There are
two
useful asymptotic behaviors for $f(y)$ in Eq.(\ref{scaling}). When
$L\ll \xi
(t)$, the whole sample evolves ``coherently'' in the sense that
$\sigma (t)$
is independent of time; hence $f(y)\sim 1$ as $y\rightarrow 0$. It
implies
that, in the limit of large $t$ ($L\ll \xi (t)$), $\sigma \sim
A_\sigma
L^\chi $. At the opposite limit, when $L\gg \xi (t)$, the local
surface
roughness has not detected the existence of the boundary of the
sample, so $
\sigma (t)$ does not depend on $L$. One then deduces that $f(y)\sim
y^{-\chi
}$ as $y\rightarrow \infty $.\cite{family} That is, in the limit of
small $t$
, $\sigma \sim A_\sigma (\tilde{\nu}t)^\beta \equiv A_tt^\beta $,
with $
\beta =\chi /z$.

In order to explain the above scaling phenomena and calculate the
relevant
exponents, a number of continuum models have been proposed.\cite
{wv,lai,Villain91,kpz} In these models, $H({\bf r},t)$ is
coarse-grained.
The interface growth is modeled by noise-driven dynamics
\begin{equation}
\frac{\partial h}{\partial t}={\rm F}[h]+\eta ({\bf r},t).
\label{heightequation}
\end{equation}
Here $h=H-\langle H\rangle ,\langle H\rangle $ being the average
height and $
\eta ({\bf r},t)$, representing fluctuations in the deposition flux,
is
modeled by a Gaussian white noise with the two-point correlation
function
given by
\begin{equation}
\langle \eta ({\bf r_1,}t_1)\eta ({\bf r_2},t_2)\rangle =2\eta
_0\delta (
{\bf r_1}-{\bf r_2})\delta (t_1-t_2).  \label{noise}
\end{equation}
The rest of the growth dynamics is lumped into ${\rm F}[h]$ in which
different driving forces for adatom movement are represented by
different
terms. A characteristic of these models is that the resulting
surfaces are
usually driven towards being self-affine, resulting in the above
scaling
behaviors.

The continuum models can be classified into two classes: conservative
and
non-conservative. In non-conservative dynamics, the flow of atoms
onto the
surface is assumed to be normal to the surface. The
Kardar-Parisi-Zhang
(KPZ) model, in which ${\rm F[h]}=\nu \nabla ^2h+\lambda _1({\bf
\nabla }
h)^2 $ with $\nu >0$, represents the lowest-order realization of such
dynamics. Here $\lambda _1({\bf \nabla }h)^2$ accounts the fact that
the
growth is normal to the interface, and 
the desorption, accounted by $\nu
\nabla
^2h$, is assumed to be important.\cite{Villain91,kpz} The exponents
of this
model for dimensionality ($d$) of $1+1$ are known exactly: $\chi
=1/2$ and $
z=3/2$, while only numerical results are available for
$d=2+1$.\cite{kpzd2}

In conservative dynamics, one only includes the flow of atoms
parallel to
the surface. Therefore, ${\rm F[h]}$ must take the form
\[
{\rm F}[h]=-{\bf \nabla \cdot J,}
\]
where J is the surface adatom current and the total volume $\int \int
dx\,dy\,h(x,y)$ is conserved. It has been argued that conservative
dynamics
is the main scenario that occurs in the molecular beam epitaxy (MBE)
growth. In
particular, surface diffusion, rather than desorption, is the
dominating
factor.\cite{wv,lai,Villain91} The surface diffusion is assumed to be
driven
by the energetics of adatoms on surfaces so that ${\bf J}$ obeys the
Fick's
law ${\bf J}=-\alpha {\bf \nabla }\mu $, where $\mu $ is the chemical
potential on surfaces. A couple mechanisms contributing to $\mu $
have been
proposed. First, because adatoms have less (more) bonding
opportunities when
they reside on ``mountains'' (``valleys''), $\mu $ is proportional to
the
curvature.\cite{herring51} Secondly, adatoms on slopes either have
higher
kinetic energy or their chemical bonds are more stretched. One thus
expects
that they have higher $\mu $. This effect may be accounted by the
term $
(\nabla h)^2.$\cite{lai} Combining these two mechanisms, Lai and Das
Sarma
(LS) investigated the model in which ${\rm F[h]}=-K\nabla ^4h+\lambda
_2\nabla ^2({\bf \nabla }h)^2.$\cite{lai} Here $-K\nabla ^4h$ is the
lowest-order contribution from the curvature. They found the
exponents to be
$\chi =2/3$ and $z=10/3$ for $d=2+1$.

The existence of a Laplacian term is an important question, for in
order to
see the exponents predicted by the LS model, there must be {\em no
Laplacian
term}. A non-vanishing Laplacian term, no matter how small, will make
$
\lambda _2$ irrelevant, resulting in different scaling exponents from
the LS
results.\cite{Tang91} There are several mechanisms that can generate
Laplacian terms. The typical way to generate the Laplacian term is
via the
desorption, which is small but is not identically
zero.\cite{barabasi} If
surfaces are liquid-like, a Laplacian term can arise from
non-vanishing
surface tension. More recently, Villain\cite{Villain91} argued that
the
combination of surface diffusion on terraces and step-edge
(Schwoebel)
barriers results in a Laplacian term. The coefficient $\nu $ of the
Laplacian term, however, is negative when growing on singular (i.e.,
high-symmetry) surfaces.

Theoretically, a negative $\nu $ leads to unstable surface growth,
producing
large-scale features on surfaces. This phenomenon was observed in a
recent
Monte Carlo study.\cite{johnson94} For continuum models, when the
negative
Laplacian term is combined with the surface diffusion term $-K\nabla
^4h$
and the KPZ non-linear term, the resulting dynamics (without noise),
known
as the Kuramoto-Sivashinsky (KS) equation, is particularly
interesting. \cite
{KSeq} The KS equation has been used to describe a wide class of
phenomena
associated with instabilities such as flame propagation and chemical
turbulence. It is linearly unstable, but is non-linearly chaotic,
exhibiting
spatio-temporal chaos. Many years ago, Yakhot\cite{Yakhot81}
conjectured
that the non-linear term in the KS equation (with random initial
conditions)
will ``renormalize'' the negative $\nu $ so that the effective
large-scale
behavior of the KS equation can be described by the KPZ equation with
a
positive, effective $\nu $. In one dimension, this is confirmed by a
recent
work of Chow and Hwa.\cite{chow95} In two dimensions, however,
conflicting
results have been reported.\cite{Lvov,Jay} There appears to be no
consensus
about the scaling behavior of the KS equation at large scales, though
they
all agree that an effective, positive $\nu $ must be present at large
length
scales. Experimentally, quantitative evidence of a non-vanishing
Laplacian
term was not established until recently,\cite{mou96} although an
earlier
study used a negative $\nu $ to explain the large-scale features in
homo-growth of GaAs films.\cite{johnson94}

A negative $\nu $ is not the only possible mechanism for generating
large
features on surfaces, i.e. rough surfaces. In fact, it has been
recognized
that a flat surface under stress may become unstable or
metastable.\cite
{tersoff} Several stress-induced instability mechanisms have been
proposed.
In the continuum elastic theory,\cite{spencer} an elastic energy is
added to
the chemical potential, resulting in a linear term, $ck^3h({\bf
k},t)$, in
Fourier space. Here $c$ is a positive constant. Tersoff et al. have
proposed
a long-range attractive interaction between steps on vicinal strained
layers.
\cite{tersoff} The interaction leads to step-bunching instability,
and
effectively, it also introduces the term {\bf \ $k^3h({\bf 
k},t)$ }but with a
different coefficient{\bf \ $c^{\prime } $.} It is important to note
that
these theories were derived for strained films without dislocations.
The $k^3
$ term is nonlocal and not analytic in real space, and thus is absent
in the
gradient expansion of ${\rm F}[h]$.

\section{Experiments and Results}

GeSi films studied here were grown either by MBE or by low pressure
chemical
vapor deposition (CVD).\cite{cvd} In order to minimize threading
dislocation
density, the Ge concentrations in these films were increased linearly
with
the average film thickness until desired compositions were achieved.
The
growth temperature for these samples was sufficiently high so as to
achieve
strain relaxation during growth.\cite{fitzgerald} The average lattice
constants of the films are the same as those of bulk crystals, i.e.,
completely relaxed. Detailed characterization of these samples is
given in
Refs. 6 and 26. We concentrate on samples with approximately the same
final
Ge composition but grown under different conditions. We also examine
samples
grown with different grading rates. Relevant growth information and
non-universal amplitudes obtained from our analysis for five samples
are
summarized in Table I.

The surface roughness was measured using a scanning force microscope
(SFM).
Large-scale (48 $\mu {\rm m})^2$ SFM images of all five samples are
shown in
Figs. 1 (a)-(e). All samples except sample C display a long-range
ordered
cross-hatch pattern on the surface.\cite{notripple} We shall see that
the
ordering of the cross-hatch pattern does not affect the scaling
behavior of
the surface roughness. In Figs. 2(a) and 2(b), we show samples B and
C
respectively, on a magnified scale, (14.5 $\mu {\rm m})^2$. The line
cuts
(height changes versus lateral distances) indicated on Figs. 2(a) and
2(b)
are shown in Figs. 2(c) and 2(d), respectively. To determine the
surface
roughness at length scales varying from 1 nm to 100 $\mu $m, images
(256
pixel $\times $ 256 pixel) similar to Figs. 1 (a)-(e) were taken at
different length scales and on at least two random spots for a given
sample.
When performing scaling analysis, we divided each image into smaller
images (
$128\times 128$, $64\times 64$, ..., $2\times 2$). The roughness for
a given
length scale $L$ was obtained by first calculating the rms roughness
inside
each $L\times L$ image, then averaging over the ensemble of images of
the
same size.\cite{rao} To obtain data for $L$ from $1$ nm to $100\,\mu
$m, we
combine results from several images of different sizes.

Fig. 3(a) shows $\sigma $ versus $L$ in a log-log plot for all five
samples.
These curves all show similar behaviors, i.e., $\sigma \propto L^1$
at small
length scales over {\it 3 decades} in both $\sigma $ and $L$, and
$\sigma $
flattens out with zero slopes ($\chi \sim 0$) above some
sample-dependent
crossover length $L_\nu =0.5\sim 10\mu $m. At larger length scales,
all
surfaces are flat in the sense that the surface roughness reaches a
saturated value ($\sigma _0$) that is almost independent of the
system size.
For the five samples we studied, $\sigma _0$ ranges from 37 to 186
\AA , and
$L_\nu $ from 0.7 to 5.5 $\mu $m (see Table I). These differences are
more
easily seen in Figs. 2(a)-(d), especially the difference for $\sigma
_0$
[Figs. 2(c) and 2(d)]. Note that the y-axis (height change) in Fig.
2(d) is
twice that in Fig. 2(c). When we re-scale $L$ by $L_\nu $ and $\sigma
$ by $
\sigma _0$, all the data collapse onto one universal curve, as shown
in Fig.
3(b). The successful collapsing demonstrates that, despite the
differences
in synthesis parameters and apparent surface morphology, all samples
belong
to the same growth universality class governed by $\sigma /\sigma
_0=f(L/L_\nu ) $, where the collapsed data points in Fig. 3(b) trace
out the
universal scaling function $f$. The collapsing of data indicates that
all
samples can be described by the same continuum model, if there exists
an
appropriate one. It also implies that the order of the cross-hatch
pattern
is not relevant since the surface of Sample C does not display the
ordered
the cross-hatch pattern but can still be collapsed with the other
data.

A simple power counting suggests that scaling results shown in Fig. 3
can be
accounted by the linear model:
\begin{equation}
{\rm F[h,t]}=\nu \nabla ^2h-K\nabla ^4h.  \label{exptF}
\end{equation}
Since the $-K\nabla ^4h$ term represents surface diffusion\cite
{Villain91,mullins57} which tends to stabilize growth, $K$ is taken
to be
positive. In this model ($d=2+1$), when $-K\nabla ^4h$ dominates,
$\chi =1$;
while if $\nu \nabla ^2h$ dominates, $\chi =0$.\cite{barabasi} The
crossover
length $L_\nu $ between the two scaling regimes is set by $2\pi
(K/|\nu
|)^{1/2}$. The good agreement in scaling exponents between
experimental
results and the model shows that surface diffusion dominates at small
length
scales.

In a more detailed analysis, we introduce an infrared cutoff
$k_0$($\equiv
2\pi /L$) and an ultraviolet cutoff $\Lambda $ ($\equiv 2\pi /a$,
where $a$
is the lattice constant)\cite{edwards} so that the roughness can be
calculated via the integral
\begin{equation}
\sigma (L)=\int_{k_0}^\Lambda \frac{d^2{\bf k}}{(2\pi )^2}\int
d\omega {\bf
\,\langle }h({\bf k,\omega })\,h(-{\bf k,}-{\bf \omega })\rangle .
\label{sigma}
\end{equation}
When $\nu \nabla ^2h$ dominates, we find $\sigma _\nu \sim \sqrt{\eta
_0/(2\pi |\nu |)}\,\sqrt{\ln (L/a)}$. In Fig. 4, we show our fit for
sample
B. The lattice constant read off from this fitting is about 6${\rm
\AA }$,
which is comparable to the lattice constant of Ge$_0._3$Si$_0._7$.
Similar
fittings have also been done for the other samples, and they are all
consistent with this form. In this regime, the non-universal
amplitude $
A_\sigma ^{(\nu )}$ is $\sqrt{\eta _0/(2\pi |\nu |)}$. The
experimentally
observed values for $A_\sigma ^{(\nu )}$ ranges from 12 to 65 \AA
(see Table
I). The saturation roughness $\sigma _0$ is related to $A_\sigma
^{(\nu )}$
via the relation: $\sigma _0=A_\sigma ^{(\nu )}\sqrt{\ln (L_0/a)}$,
where $
L_0$ is the scan size in the Laplacian term dominant region. Since
for all
samples$\sqrt{\ln (L_0/a)}$ is about ${\rm 3}$, $\sigma _0$ and
$A_\sigma
^{(\nu )}$ are of the same order. Using $L_\nu $, determined from the
crossover of the scaling exponents, and $A_\sigma ^{(\nu )}$,
determined
from a logarithmic fit of the saturation region, to calculate
$A_\sigma
^{(\nu )}\sqrt{\ln (L_\nu /a)}$, the results agree very well with
experimentally measured $\sigma _0$ for all five samples (see Table
I). The
reason $A_\sigma ^{(\nu )}\sqrt{\ln (L_\nu /a)}$ is consistently
smaller
than $\sigma _0$ comes from{\bf \ }the fact that{\bf \ $L_\nu <L_0$.}
At
small length scales ($L<L_\nu $), if $-K\nabla ^4h$ dominates, we get
$
\sigma _K\sim \sqrt{\eta _0/(16\pi ^3K)}\,L$. Here $A_\sigma ^{(K)}$
is $
\sqrt{\eta _0/(16\pi ^3K)}$, which is the slope $d\sigma /dL$ for
$L<L_\nu $
. Experimentally, we can independently determine 3 parameters:
$\sigma _0$, $
A_\sigma ^{(K)}$ and $L_\nu $. For our data to be consistent with
Eq.(\ref
{exptF}), these parameters must satisfy $\sigma _0/A_\sigma
^{(K)}\approx
4.75L_\nu $. In Fig. 5, we plot $\sigma _0/A_\sigma ^{(K)}$ versus
$L_\nu $.
We see that the linearity between $\sigma _0/A_\sigma ^{(K)}$ and
$L_\nu $
is quite good, though the slope is 1 and not 4.75. This shows that as
far as
scaling is concerned, the above linear model describes the growth of
this
real, complex hetero-growth system reasonably well.

The discrepancy between the measured slope for $\sigma _0/A_\sigma
^{(K)}$
versus $L_\nu $ and the predicted value from scaling suggests that
non-linear terms do not identically vanish. Therefore, as a test of
linearity, we also examine the value of $\theta \equiv |\langle
h^3\rangle
|^{1/3}$. In all samples, we found nonzero $\theta $s (see Table I).
There
is no apparent enhancement of $\theta $ from increasing the grading
rate
(samples B and C). In Fig. 6, we show the plot of $\theta $ versus
$L$ for
sample B. Even though the data are more noisy, it shows that $\theta
$ has a
similar dependence on $L$ as that of the roughness. This indicates
that
these surfaces are not multi-affine.\cite{barabasi} The non-vanishing
of $
\theta $ implies that the up-down symmetry is broken and thus a
non-linear
term must be present. The lowest-order non-linear term appears to
have no
effect on the scaling behavior of $\sigma $ below {\rm 100 }$\mu {\rm
m}$
(including the linear relation between $\sigma _0/A_\sigma ^{(K)}$
and $
L_\nu )$, but they are important in determining factors besides
scalings
such as the slope between $\sigma _0/A_\sigma ^{(K)}$ and $L_\nu $.

We can further estimate the magnitudes of $\nu $, $K$, and $\eta _0$.
Assuming that surface diffusion is induced by variations of the
chemical
potential, $K$ is given by ${D_s\gamma \Omega
^2n}/{k_BT}$,\cite{mullins57}
where $D_s$ is the surface diffusion constant, $n$ is the areal
density of
adatoms,\cite{Villain91} $\gamma $ is the surface tension and $\Omega
$ is
atomic volume. At relevant growth temperatures, $D_s$ is of order
$10^{-5}
{\rm {cm}^2/{sec}}$,\cite{diffu} while $\gamma \sim 10^3$ erg/${\rm
{cm}^2}$
.\cite{zangwill} The values of the remaining parameters are
standard.\cite
{constant} Altogether, we find that $K\sim {\rm 10}^{-20}{\rm
{cm}^4{sec}
^{-1}}$ , which implies that $|\nu |$ is of order ${\rm 10}^{-2}{\mu
{\rm {m}
}}^2{\rm {sec}^{-1}}$ via $2\pi (K/|\nu |)^{1/2}\sim L_\nu \sim 1\mu
{\rm {m}
}$. From the relation $A_\sigma ^{(\nu )}\sim \sqrt{\eta _0/(2\pi
|\nu |)}$
and the experimental values of $A_\sigma ^{(\nu )}$ and $|\nu |$, we
find $
\eta _0\sim {\rm 10}^{-23}{\rm cm}^4/\sec $.

Another important feature in Fig.~3 is that samples with faster
grading rate
(80\%Ge${\rm /}\mu {\rm {m}}$), such as samples B and C, are rougher
(with
larger $\sigma $ and $A_\sigma ^{(K)}$) for $L<L_\nu $. In addition,
$\sigma
$ for samples A, D, and E, all of which were grown at 10\%Ge$/\mu
{\rm {m}}$
grading rate, are identical at small length scales. For a relaxed
film,
strain fields are not uniform in the films, but concentrate near
dislocations.\cite{fitzgerald} The larger grading rate means that the
growth
surface is closer to the dislocations and therefore the surface
strain
fields are larger. Hence, the larger $A_\sigma ^{(K)}$s observed in
the
80\%Ge$/\mu {\rm {m}}$ grading rate samples suggest that strain
enhances surface roughness. From a different point of view, since the
primary effect of the $-K\nabla ^4h$ term is to decrease roughness,
our
results indicate that strain fields suppresses adatom diffusion on
the
surface.

\section{Discussion and conclusions}

We now examine our results more closely. First, the relation
$A_\sigma
^{(\nu )}/A_\sigma ^{(K)}\approx L_\nu $ has a natural geometrical
meaning.
It reflects the wavy nature of the surface morphology in our samples:
if one
treats $A_\sigma ^{(\nu )}$ approximately as the amplitude of the
wave, $
A_\sigma ^{(K)}$ as the slope from the valley to the peak, and $L_\nu
$ as
the width from peak to peak, the relation $A_\sigma ^{(\nu
)}/A_\sigma
^{(K)}\approx L_\nu $ follows from the definition of slope. In
Appendix A,
we calculate the surface roughness of a continuum sinusoidal surface.
The
surface roughness of this model surface has a similar saturation for
the $
L\gg L_\nu $, but in the opposite limit $L\ll L_\nu $ the roughness
does not
have exact linear scaling relation with $L$. One should be further
cautious
that the real surface is not exactly a sinusoidal wave. In Figs. 2(c)
and
2(d), we show the surface roughness line cuts for samples B and C,
respectively. The local slopes vary from $<10$ \AA /$\mu $m to $>800$
\AA /$
\mu $m for sample B and $>1000$ \AA /$\mu $m for sample C, indicating
that
these surfaces contain more than one wavelength.

Secondly, the experimentally observed $\nu $ ($\equiv \nu _E$) is an
effective one and must be positive. If $\nu $ is negative, $h({\bf
k},t)$
grows exponentially. The roughness would not saturate at large times.
In
this case, one finds the roughness $\sigma (t,L)$ by expressing the
height
in terms of $\eta $
\[
h({\bf k},t)=\int_0^tdt^{\prime }\exp (|\nu |k^2t^{\prime })\,\eta
({\bf k}
,t^{\prime }),
\]
and calculates the correlation function $\langle h({\bf k},t)h({\bf
k}
^{\prime },t)\rangle $. The roughness is then determined by the
integral $
\int d^2{\bf k}\int d^2{\bf k}^{\prime }\langle h({\bf k},t)h({\bf k}
^{\prime },t)\rangle $. We obtain
\begin{eqnarray}
\sigma ^2(L,t) &=&\frac{\eta _0}{2\pi |\nu |\,}\left[ \int_{2\pi
/L}^{2\pi
/a}\frac{\exp (2|\nu |k^2t)}kdk-\ln \frac La\right]   \nonumber \\
&=&\frac{\eta _0}{4\pi |\nu |\,}\left[
{\rm Ei}
\left( 8|\nu |\pi ^2\frac t{a^2}\right) -
{\rm Ei}
\left( 8|\nu |\pi ^2\frac t{L^2}\right) -2\ln \frac La\right]
\label{sigmaT}
\end{eqnarray}
where $
{\rm Ei}
$ is the exponential integral function and we have assumed that
initially
the surface is flat, i.e., $\sigma ^2(L,0)=0$. For our samples, the
growth
time is about ${\rm 10}^4$ sec and $\Lambda \simeq {\rm 1\AA
}^{-1}\approx
10^4(2\pi /L)$, so the order of $2|\nu _E|\Lambda ^2t$ is ${\rm
10}^{10}$,
which is very large. As a result, the first term dominates so that
$\sigma
(L,t)$ is approximately independent of $L$. Therefore, if $\nu _E$ is
negative, its value, when combined with Eq.(\ref{sigmaT}) and the
observed $
\sigma _0$, yields a value of $\eta _0$ at the order of ${\rm
10}^{-10^9}
{\rm {cm}^4{sec}^{-1}}$, which is unreasonably small. It therefore
implies
that $\nu _E$ can not be negative.

What is the bare growth equation that can generate the observed
cross-hatch
pattern and yet result in a positive $\nu _E$ at large length scales?
At
first, it seems that the term $ck^3h({\bf k},t)$, induced by the
stress, must
be present. However, as mentioned earlier, this term was derived only
for
strained films without dislocations, while our samples are completely
relaxed. Therefore, we do not include it and consider the case when
the bare
$\nu $ ($\equiv \nu _0$) that appears in the growth equation is
negative.
\cite{Villain91} This coefficient $\nu _0$ enters the roughness
$\sigma (L,t)
$ at very early times. When $t$ is very small, because the height $h$
of the
surface is still small, the non-linear terms, which are higher order
in $h$,
can be neglected. Therefore, ${\rm F[h]}$ in the growth equation can
be
simply approximated by a negative Laplacian term $\nu _0\nabla ^2h$.
Eq.(\ref
{sigmaT}), with $\nu $ replaced by $\nu _0$, then describes the
roughness
only when $t$ is very small. The value of $\nu _0$ may be obtained,
for
example, by measuring the first and second time derivatives of
surface
roughness via the relation
\begin{equation}
\left. \frac{d^2}{dt^2}\sigma ^2(L,t)\right| _{t=0}=|\nu _0|\left[
\Lambda
^2+(2\pi /L)^2\right] \left. \frac d{dt}\sigma ^2(L,t)\right|
_{t=0}\,,
\label{sigmadt}
\end{equation}
where we have expressed $\eta _0$ in terms of the initial increase
rate of $
\sigma ^2(L,t)$. After a short period of initial growth, the height
$h$ has
grown so large that one cannot ignore the effects of non-linearity
any more.
According to the standard picture of pattern formation,\cite{pattern}
non-linear terms then saturate the initial unstable growth, resulting
in the
final morphology. If $\nu _0$ is negative due to Schwoebel barriers,
\cite
{Villain91} we expect the growth on vicinal surfaces will be much
smoother
because the bare $\nu $ is positive even at early times. This is
indeed
observed experimentally.\cite{hsu}

There are a couple of important issues that need to be addressed: (1)
What
are the large-scale\cite{note1} scaling behaviors of the final
morphology?
and (2) what are the non-linear terms that enter the growth equation?
{}From
our results, it is clear that up to ${\rm 100\mu m}$, 
the scaling behaviors at large
length scales are captured by the growth equation composed of a
Laplacian
term with an effective, {\it positive}{\bf \ }$\nu $. This fact also
provides us some insight about the leading order non-linear term. In
Appendix B, we examine in {\bf \ }detail the combined growth
equation, ${\rm
F[h,t]}=\nu _0\nabla ^2h-K\nabla ^4h+$ $\lambda _2\nabla ^2({\bf
\nabla }h)^2
$, within the framework of Dyson-Wyld renormalized perturbation
theory.\cite
{Lvov} We show that $\nu _0$ does not get any correction that comes
from $
\lambda _2\nabla ^2({\bf \nabla }h)^2$ in the large length scale
limit. This
results reflects what we mentioned earlier: $\lambda _2$ is
irrelevant in
the presence of a Laplacian term. Thus, we are left with $\lambda
_1({\bf
\nabla }h)^2$ as the only possible leading non-linear term. The
resulting
growth equation is the KS equation, driven by a Gaussian noise. As
discussed
in Section II, the situation of theoretical work on the large length
scale
scaling behaviors of the KS equation\cite{KSeq,Yakhot81,chow95} is
not clear
now. In fact, the reported results for the two-dimensional case are
conflicting. Nevertheless, they all agree that an effective and
positive $
\nu $ must present at large length scales. This appears to be
precisely what
we observe. Although we are not able to resolve the theoretical
conflict
using these experimental results, detailed analyses of our results
set a
minimum size of the sample for resolving this conflict. Because we
did not
observe the scaling exponents predicted by the KPZ equation, if
$\lambda
_1(\nabla h)^2$ survives in the effective equation, its effect must
be small
for length scale below ${\rm 100\mu m}$.{\bf \ }The lower end of this
non-linear term dominant regime must match with the upper end of the
regime
dominated by the Laplacian term. Since the characteristic timescale
in the
non-linear regime is $\tilde{\nu}L^z$, matching the timescales in two
regimes defines a crossover length $L_c$
\begin{equation}
L_c\approx \left( \frac \nu {\tilde{\nu}}\right) ^{1/(2-z)},
\label{L_c}
\end{equation}
where $z$ and $\tilde{\nu}$ are the corresponding exponent and
non-universal
amplitude in the KPZ regime. Another relation can be obtained by
matching
the roughness. We find that
\begin{equation}
A_\sigma \gtrsim \frac{A_\sigma ^{(\nu )}\ln (L_c/a)}{L_c^\chi },
\label{amplitude}
\end{equation}
where $A_\sigma $ and $A_t$ (in the following equation) represent the
corresponding non-universal amplitudes in the KPZ regime. Since $
A_t=A_\sigma (\tilde{\nu})^\beta $, we obtain
\begin{equation}
L_c \gtrsim \sqrt{\nu \,\left[ A_\sigma ^{(\nu )}\ln
(L_c/a)/A_t\right]
^{z/\chi }}.  \label{L_c1}
\end{equation}
Because we do not observe any crossover below ${\rm 100 \mu m}$,
$L_c$ has to be larger than ${\rm 100 \mu m}$. In addition, by setting
$L_c$ on the right hand side of Eq.(\ref{L_c1}), we can get an estimate
of the lower bound for $L_c$. 
Using the following figures: $z/\chi \approx 4$, $A_\sigma ^{(\nu )}\ln
(L_0/a)\approx {\rm 10^2 \AA }$ for $L_0\approx {\rm 100 \mu m}$, 
$A_t \sim 1 \AA / ({\rm sec})^{\beta}$ and 
$\nu
\approx 10^{-2}{\mu {\rm {m}}}^2{\rm {sec}^{-1},}$ we find that 
the lower bound is
about 1mm, which is beyond our measurement range.

In conclusion, we have observed a universal scaling behavior for the
surface
morphology of compositionally-graded, relaxed GeSi/Si(001) films.
Quantitative analyses on scaling exponents and non-universal
amplitudes show
that the scaling behaviors for samples grown under different
conditions all
belong to the same universality class, which can be described by the
linear
model $F[h,t]=\nu \nabla ^2h-K\nabla ^4h$
for $ {\rm 1 nm} \lesssim {\rm L}  \lesssim {\rm 100} \mu $ m.
In combination with further
theoretical analyses, it is argued that the underlying growth model
is the
KS equation driven by a Gaussian noise. Our results indicate that as
far as
the roughening exponents are concerned, up to 100 $\mu$m, the
effective
theory of the KS equation is the linear model, pushing the length
scale for
observing the KPZ scaling exponents to be above 1 mm.

\section{Acknowledgments}

C.-Y. Mou gratefully acknowledges support from the National Science
Council
of the Republic of China under Grants No. NSC 86-2112-M-007-006. J.
W. P.
Hsu acknowledges the support of a Sloan Foundation Fellowship. Work
done at
the Univ. of Virginia is partially funded by NSF DMR-9357444.
\newpage\

\appendix

\section{}

In this appendix, we demonstrate some features of the surface
roughness for
a continuum wavy surface. Specifically, we shall consider a
sinusoidal
surface $y=y_0\sin (2\pi x/L_0)$, but many features do not depend on
the
particular form we choose. Since one only measures discrete points in
experiments (the number of pixels determining the distance between
adjacent
points), we assume that $L_0=N_0\varepsilon $, where $N_0$ is a
positive
integer and $\varepsilon $ is the distance between adjacent points.
The
experimental data then consist $2N+1$ equidistant points ($x_n${\bf ,
}$h_n$
) centered at ($x_0$, $h_0$), where we have defined $h_i=y_0\sin
(2\pi
x_i/L_0)$ and $x_n-x_0=n\varepsilon $ with $n$ ranging from $-N$ to
$N$. The
surface roughness can be found by evaluating the following sums
\begin{eqnarray}
\langle h\rangle &=&\frac{y_0}{2N+1}\sum_{n=-N}^{n=N}\sin \frac{2\pi
(n\varepsilon +x_0)}{N_0\varepsilon },  \label{sum1} \\
\langle h^2\rangle &=&\frac{y_0^2}{2N+1}\sum_{n=-N}^{n=N}\,\left[
\sin \frac{
2\pi (n\varepsilon +x_0)}{N_0\varepsilon }\right] ^2.  \label{sum2}
\end{eqnarray}
For large $N_0$, these sums become integrals. Evaluating these
integrals,
one obtains a surface roughness $\sigma (N,L_0,x_0)$, which depends
on both $
x_0$ and $L_0$. If the modulation on the surface contains only one
wavelength, the global surface roughness is simply an average of
$x_0$ over
the period $L_0$. We find that
\begin{equation}
\sigma =y_0\left[ \frac L{2L+\varepsilon }-\left(
\frac{L_0}{2L+\varepsilon }
\frac 1{2\pi }\right) ^2\left( 1-\cos \frac{4\pi L}{L_0}\right)
\right]
^{1/2},  \label{sigmasin}
\end{equation}
where $L${\ (=$N\varepsilon $) is the size of the sample. In this
case, $
\sigma $ approaches $y_0$ as $L$ approaches $\infty $. For $L<L_0$,
$\sigma $
does not scale; instead, it oscillates with $L$. }

\section{}

In this appendix, we shall investigate the large scale behavior of
the model
\begin{equation}
\frac{\partial h}{\partial t}=\nu _0\nabla ^2h-K\nabla ^4h+\lambda
_2\nabla
^2\left( \nabla h\right) ^2+\eta ,  \label{eq_appendix}
\end{equation}
where $\nu _0$ is {\it negative}. Specifically, we shall show that
there is
no correction to $\nu _0$ at large length scales.

Two important functions we shall work with are the response function
${\rm G}
(k,\omega )$ and the two-point correlation function ${\rm U}(k,\omega
)$.
They are defined by
\begin{eqnarray}
{\rm G}(k,\omega ) &\equiv &\left\langle \frac{\delta h({\bf
k},\omega )}{
\delta \eta ({\bf k},\omega )}\right\rangle _{\eta \rightarrow 0}\,,
\label{G} \\
{\rm U}(k,\omega ) &\equiv &\left\langle h({\bf k},\omega )\,h(-{\bf
k}
,-\omega )\right\rangle .  \label{U}
\end{eqnarray}
The generic forms of the these function are\cite{Lvov}
\begin{eqnarray}
{\rm G}(k,\omega ) &=&\frac 1{-i\omega +\gamma _k+\Sigma (k,\omega
)}\,,
\label{selfenergy} \\
{\rm U}(k,\omega ) &=&|{\rm G}(k,\omega )|^2\left[ 2\eta _0+\Phi
(k,\omega
)\right]
\end{eqnarray}
where $\gamma _k=-\nu _0k^2-Kk^4$, $\Sigma (k,\omega )$ and $\Phi
(k,\omega
) $ are the self-energies. We shall be interested in scaling
solutions in
which ${\rm G}(k,\omega )=g(\omega /\nu k^z)/vk^z$ and ${\rm
U}(k,\omega
)=u(\omega /\nu k^z)/k^\Delta $. These scaling solutions are
asymptotically
correct only for $k\rightarrow 0$ and $\omega \rightarrow 0$.
Therefore, $
\gamma _k$ has to be subdominant to $\nu k^z$ for $k\rightarrow 0$;
hence $
z\leq 2$. The exponents $z$ and $\Delta $ are related to $\chi $ via
the
relation: $2\chi =\Delta -d-z$. We shall show that $z=2$ is not
acceptable.

Let us first consider the one-loop diagrams in the renormalized
Wyld-Dyson
perturbation theory.\cite{Lvov} The self-energy is
\begin{equation}
\Sigma ^{(2)}(k,\omega )=4(\lambda _2)^2\int_{{\bf q}}\int_\Omega
\left[
k^2( {\bf k}-{\bf q})\cdot {\bf q}\right] \left[ q^2{\bf k}\cdot
({\bf k}-
{\bf q} )\right] {\rm G}(q,\Omega ){\rm U}(|{\bf k}-{\bf q}|,\omega
-\Omega
).  \label{phi2}
\end{equation}
By appropriate changes of variables: ${\bf Q}={\bf q}/k$, $t=\Omega
/\nu q^z$
, and $s=\omega /\nu k^z$, we can rewrite Eq. (\ref{phi2}) in the
following
forms
\begin{equation}
\Sigma ^{(2)}(k,\omega )=k^{d+8-\Delta }\,\sigma ^{(2)}(s,\Lambda
/k,m_0/k).
\label{phi2_1}
\end{equation}
Here $\sigma ^{(2)}$ is a function of $s$, $\Lambda /k$, and $m_0/k$.
$
\Lambda $ and $m_0$ are the ultra-violet (UV) and the infrared (IR)
cutoffs
in the $q$-integrals. These forms imply that if the $q$-integrals are
divergent, the leading terms in $\Sigma ^{(2)}(k,\omega )$ must take
the
form $k^{d+8-\Delta -\delta }\Lambda ^\delta $ or $k^{d+8-\Delta
+\delta
^{\prime }}m_0^{-\delta ^{\prime }}$, depending on whether the
divergences
are UV or IR. Here both $\delta $ and $\delta ^{\prime }$ must be
positive.
If the divergences are UV type, we set all internal momentum, such as
$q$
and $|{\bf k}-{\bf q}|$, to $\Lambda $, and all internal frequencies
to $
\Lambda ^z$. Only external momentum are left intact. Since in $\Sigma
^{(2)}(k,\omega )$ the first vertex carries one $k^2$ and the final
vertex
carries one external momenta ${\bf k}$, the lowest term would be
$O(k^2{\bf
k })$. However, because $\Sigma (k,\omega )$ depends only on $k$,
this term
must vanish. Hence, the leading term is $k^4\Lambda ^{d+4-\Delta }$.
The
sub-leading terms are $k^6\Lambda ^{d+2-\Delta }$, $k^8\Lambda
^{d-\Delta }$
, and so on, with the power of $\Lambda $ decreasing until the power
of $
\Lambda $ becomes negative. For the above to be correct, one requires
$
\Delta \leq d+4$, and the difference between $\Delta $ and $d+4$
decides the
number of sub-leading terms. It is clear that the UV divergences do
not
contribute any correction to $\nu _0$.

On the other hand, if the divergences are IR type, one sets $q$ or
$|{\bf k}
- {\bf q}|$, but not both, to $m_0$. Simple power counting leads to
the
conclusion that the leading term of $\Sigma ^{(2)}(k,\omega )$ is $
k^{6-z}(m_0)^{d+2-\Delta +z}$. The sub-leading terms are terms with
less
power of $k$, so they could correct $\nu _0$. The correction must
take the
form $k^2(m_0)^{d+6-\Delta }$. Obviously, it implies that $\Delta
\geq d+6$,
which results in $\chi \geq (6-z)/2\geq 2$. Since in the physical
regime, $
\chi \leq 1$,\cite{family2} this is also ruled out. Therefore, the IR
divergences do not contribute $\nu _0$ in the physical regime($\chi
\leq 1$)
either.

Finally, if the integral $\sigma ^{(2)}$ is convergent, it depends
only on $
s $ in the limits $\Lambda \rightarrow \infty $ and $m_0\rightarrow
0$. In
order that $\nu _o$ gets a correction, we require $\Sigma
^{(2)}(k,\omega
)\,\sim k^2$. Hence $d+8-\Delta =2$, i.e., $\Delta =d+6$. However,
this
values falls into the regime where the integral $\sigma ^{(2)}$ is
not
convergent, but IR divergent!

The above analysis can be easily generalized to higher order terms.
We find
that the dimension of ${\rm 2n}$th order terms in $\sum $ is
$n(d+8-\Delta
-z)+z.$ These terms can not be both convergent and at the same time
contribute $O(k^2)$ because it would imply $\Delta
=d+8-2/n+(1/n-1)z$, which
is greater than $d+6$ and thus falls into the IR regime. Thus these
terms
must be divergent. We find that the leading contribution of the UV
divergences to $\Sigma $ are of the form $k^4\sum\limits_{n=1}^\infty
a_n\Lambda ^{n(d+6-\Delta )-2}$, where $a_n$ is the contribution of
${\rm 2n}
$th order terms; hence the UV divergences do not correct $\nu _0$ at
all.
Similarly, for IR divergences, possible corrections to $\nu _0$
coming from
the ${\rm 2n}$th order terms must be of the form
$k^2m_0^{n(d+8-\Delta
-z)+z-2}$. It implies that $\Delta \geq d+8+(1/n-1)z-2/n$, and hence
$\chi
\geq 4+(1/2n-1)z-1/n\geq 2$, which is not in the physical regime, so
the IR
divergences do not contribute $\nu _0$ at all in the physical
regime($\chi
\leq 1$). We thus conclude that to all orders in the Wyld-Dyson
renormalized
perturbation expansion, there is no correction to $\nu _0$.

\clearpage

\begin{table}[h]
\caption{Detailed growth parameters ($R_1$ and $R_2$ are the growth
and
grading rates) and non-universal amplitudes obtained from our
analysis of
samples A - E : $\sigma_0$ and $\theta_0$ are saturated values of
surface
roughness and $|\langle h^3 \rangle |^{1/3}$ ,
respectively.}\vspace{0.1in}
\begin{tabular}{|c|c|c|c|c|c|}
\hline
sample & A & B & C & D & E \\ \hline
Ge(\%) & 30 & 30 & 30 & 40 & 30 \\ \hline
growth method & MBE & MBE & MBE & CVD & CVD \\ \hline
T($^{\circ}$ C) & 900 & 900 & 900 & 850 & 650 \\ \hline
P(mT) & -- & -- & -- & 2 & 50 \\ \hline
$R_1$(${\rm {\AA} / sec}$) & 3 & 3 & 3 & 7 & 12 \\ \hline
$R_2$(Ge\%$/ \mu {\rm {m}}$) & 10 & 80 & 80 & 10 & 10 \\ \hline
$L_{\nu}$($\mu {\rm {m}}$) & 1.5 & 0.67 & 1.05 & 2.0 & 5.5 \\ \hline
$\sigma_0 $ (\AA) & $37 \pm 4$ & $65 \pm 4$ & $186 \pm 15$ & $61 \pm
7$ & $
124 \pm 5$ \\ \hline
$\theta_0$(\AA) & $13 \pm 4$ & $19 \pm 5$ & $91 \pm 38$ & $29 \pm 11$
& $51
\pm 6$ \\ \hline
$A_{\sigma}^{(\nu)}$ (\AA) & 12 & 22 & 65 & 20 & 38 \\ \hline
$A_{\sigma}^{(K)}$ (10$^{-4}$) & 26 & 87 & 169 & 29 & 28 \\ \hline
$A_\sigma ^{(\nu)}\sqrt{\ln (L_{\nu}/a)}$ (\AA) & 34 & 58 & 178 & 57
& 115
\\ \hline
\end{tabular}
\end{table}

\clearpage

\begin{center}
{FIGURE CAPTIONS}
\end{center}

{\sc FIG. 1}. SFM images of samples listed in Table I: A(a), B(b),
C(c),
D(c), and E(d), respectively. The image sizes are ($48\mu $m) $\times
$ ($
48\mu $m).\\

{\sc FIG. 2}. (14.5 $\mu $m) $\times $ ( 14.5 $\mu $m) SFM images of
samples
B(a) and C(b), respectively. The surface roughness line cuts (height
change
versus lateral distance) indicated by the white lines in (a) and (b)
are
shown in (c) and (d), respectively.\\

{\sc FIG. 3}. (a) Surface roughness ($\sigma $) versus $L$ for all 5
samples
in Table I. The calculation of $\sigma (L)$ is given in the text. The
line
represents $\sigma \propto L^1$. (b) $\sigma /\sigma _0$ vs. $L/L_\nu
$ for
all samples. \\

{\sc FIG. 4}. A logarithmic fit to the data of sample B. The lattice
constant obtained from this fit is about 6${\rm \AA }$, and $A_\sigma
^{(\nu
)}=\sqrt{\eta _0/(2\pi |\nu |)}$ is 22 \AA .\\

{\sc FIG. 5.} Ratio of non-universal amplitudes $\sigma _0/A_\sigma
^{(K)}$
vs. the crossover length $L_\nu $. The line is a guide to the eye
with slope
$=1$.\\

{\sc FIG. 6.} The third moment $\theta \equiv |\langle h^3\rangle
|^{1/3}$
versus $L$ for sample B.

\clearpage


\begin{references}
\bibitem{notripple}  Even though the surface features of Sample C
resemble
rippling (surface relaxation without dislocations), it is not due to
rippling. The ripples on the rippling surfaces are in the \mbox{$<$}
100
\mbox{$>$} directions. The features on Samples C are in the
\mbox{$<$} 110
\mbox{$>$} directions, same directions as the misfit dislocations and
the
cross-hatch patterns.
\bibitem{note1}  By large-scale, we mean the scales that are larger
than the
scale of the most unstable modes. In our samples, it means the scales
that
are larger than the inter-valley distances of the cross-hatch
patterns.
\end{references}
\end{document}